\documentclass[10pt,leqno]{amsart}
\usepackage{graphicx}
\usepackage{indentfirst,csquotes}

\usepackage{amssymb,amsthm,amsmath}
\usepackage{xcolor,paralist,hyperref,fancyhdr,etoolbox}
\usepackage{titlesec}

\makeatletter
\renewcommand{\section}{%
  \@startsection{section}{1}{0pt}%
  {-1.5ex plus -.5ex minus -.2ex}%  % space before
  {1.0ex plus .2ex}%                 % space after
  {\normalfont\Large\bfseries\centering}%
}

\renewcommand{\subsection}{%
  \@startsection{subsection}{2}{0pt}%
  {-1.25ex plus -.5ex minus -.2ex}%
  {0.8ex plus .1ex}%
  {\normalfont\large\bfseries}%
}

\renewcommand{\subsubsection}{%
  \@startsection{subsubsection}{3}{0pt}%
  {-1.0ex plus -.2ex minus -.1ex}%
  {0.5ex}%
  {\normalfont\normalsize\bfseries\itshape}%
}

\renewcommand{\paragraph}{%
  \@startsection{paragraph}{4}{0pt}%
  {0.8ex plus .2ex minus .1ex}%
  {-1em}%
  {\normalfont\normalsize\bfseries}%
}

\makeatother

\hypersetup{ colorlinks=true, linkcolor=black, filecolor=black, urlcolor=black }

\usepackage{lipsum}

\begin{document}

\pagestyle{plain}

\title[Angle and time-resolved polarization change induced by Kerr effect
in amorphous and crystalline SiO$_2$]%
{Angle and time-resolved polarization change induced by Kerr effect
in amorphous and crystalline SiO$_2$}

\author{}

\begin{center}
\textbf{Pierrick Lample\textsuperscript{1,2,*},
Mateusz Weis\textsuperscript{3},
Davide Boschetto\textsuperscript{3},
St\'ephane Guizard\textsuperscript{1}}\\[6pt]

\textsuperscript{1}Universit\'e Paris-Saclay and Universit\'e Cergy, CEA, LIDYL, 91191 Gif-sur-Yvette, France\\
\textsuperscript{2}Currently at Universit\'e Paris-Saclay, CNRS, LPS, 91400 Orsay, France\\
\textsuperscript{3}Laboratoire d’Optique Appliqu\'ee, ENSTA, CNRS, \'Ecole Polytechnique, Institut Polytechnique de Paris, 91761 Palaiseau, France\\[4pt]
\textsuperscript{*}\textit{Corresponding author:} \texttt{pierrick.lample@universite-paris-saclay.fr}
\end{center}

\date{\today}

\maketitle

% (optionnel) enlever le pied de page MSC si tu ne fais pas de maths pures
% \let\thefootnote\relax
% \footnotetext{MSC2020: Primary 00A05, Secondary 00A66.}

\begin{abstract}
We measure the polarization change of a beam reflected from the surface of both crystalline $\alpha$ and amorphous SiO$_2$ samples while they are photo-excited by an intense light pulse, at intensities above the nonlinear excitation threshold yet below the damage threshold. The polarization change varies with the angle between the polarization of pump and probe light, but is found to be independent of their orientation relative to the crystal axes. This behavior differs between the reflected and transmitted beams, and can be modeled by taking into account a birefringence induced by the electric field of the pump. These polarization-change effects can be very strong, with polarization rotation exceeding $90^{\circ}$, at pump intensities well below the damage threshold. We also observe a markedly 
different behavior of the reflected beam depending on whether the material is crystalline or amorphous.
\end{abstract}

\section{Introduction}

Coherent light pulses have proven highly effective for investigating the ultrafast dynamics of matter. Furthermore, the development and widespread availability of user-friendly sub-picosecond high-power, high-repetition-rate sources are opening new perspectives for laser machining. The demand for a better knowledge of all elementary processes occurring during the interaction of intense light pulse with dense matter is thus very high.

To investigate this, different pump-probe techniques have been used: for a review see  \cite{Jrgens2024}. Most of them consist of measuring the change of amplitude or phase of a transmitted or reflected probe, giving direct indication about the change of dielectric constant. While research is now progressively reaching higher resolutions, with an increase in studies employing attosecond pulses \cite{Mitrofanov2011, Schultze2012, Schiffrin2012, Volkov2023, Gneaux2020}, much less work has been devoted to the change in polarization \cite{Gertsvolf2010, Mller2020}.

In this work, we report simultaneous measurement of time resolved polarization changes induced by a pump pulse on both the reflected and transmitted probe beam. These measurements were performed for various pump–probe polarization angles, different orientations relative to the crystal axes, and several pump intensities.

\section{Experiment description}

\subsection{Experimental setup}

\begin{figure}[htbp]
    \centering
    \includegraphics[width=11cm]{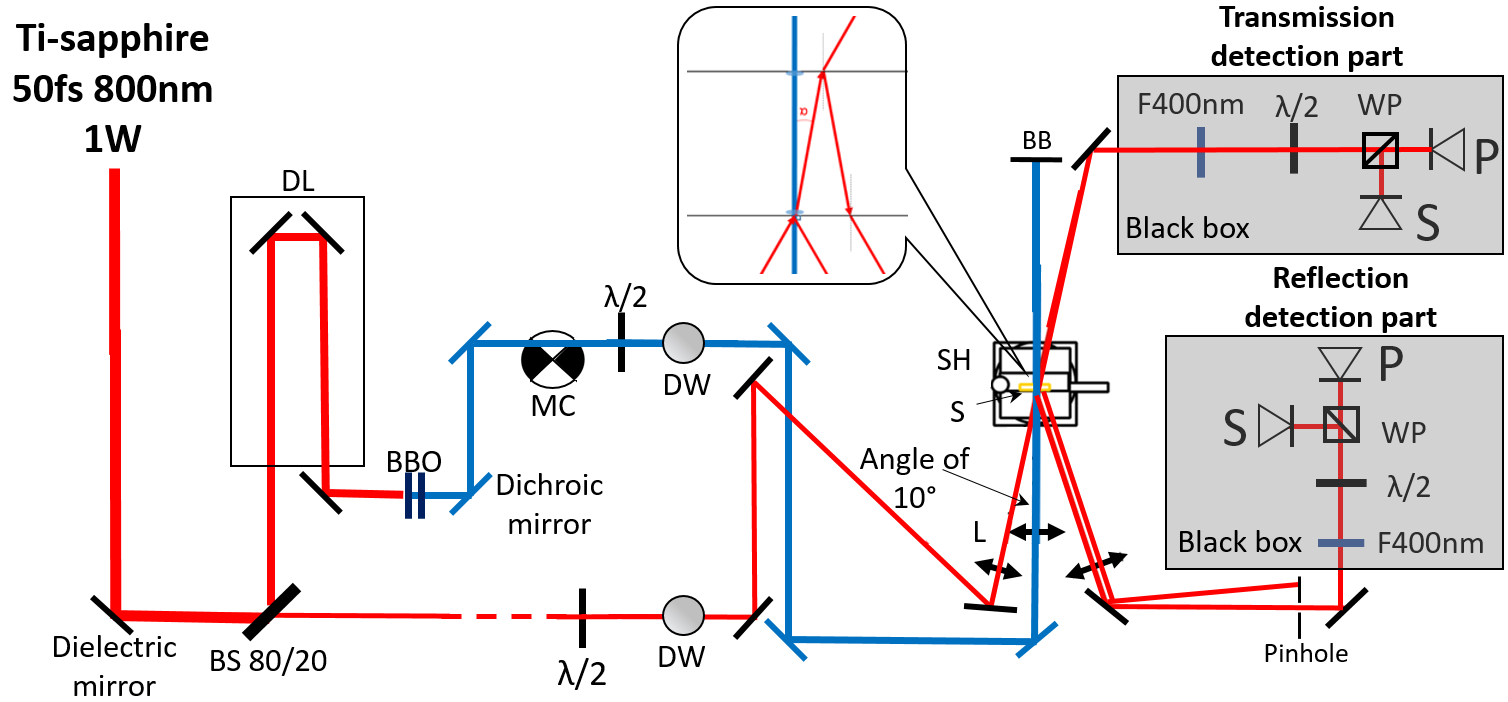}
    \caption{Schematic representation of the experimental setup. BS (beam-splitter), DL (delay line), MC (modulating chopper), DW (density wheel), L (lense), S (sample), SH (sample holder), BB (beam blocker), F400nm (high pass filter blocking 400nm), WP (Wollaston prism).}
    \label{fig1}
\end{figure}

To investigate ultrafast polarization dynamics, we use a Ti:Sa laser operating at $1$kHz with a central wavelength of $800$nm and a pulse duration of $50$fs. As shown in Figure \ref{fig1}, the beam is split into pump and probe arms. The probe remains at $800$nm, while the pump is frequency-doubled to $400$nm using a BBO crystal to enable efficient 3-photon excitation in quartz. The pump is modulated at $500$Hz, and a lock-in amplifier isolates the signal component associated with pump excitation.

Both beams reach the crystalline or amorphous $SiO_2$ sample at different angles: the pump has a normal incidence, while the probe impinges at an incidence angle of $10^{\circ}$ angle, with their relative timing controlled by a mechanical delay line. The probe beam is partially transmitted and reflected by the sample, and both components are collected using two dedicated detection setups. This oblique incidence was deliberately chosen to suppress parasitic reflections from the back surface of the sample, which arise due to the material’s transparency at $800$nm. The interaction region is imaged onto the detection part using a lens, and an iris placed before the detection scheme blocks the unwanted contribution coming from the rear surface (insert in Figure \ref{fig1}). The pump and probe beams are both linearly polarized: the probe is $P$, while the pump has its electric field oriented at an angle $\theta$ within the $P$/$S$ probe's polarization plane. $P$ and $S$-polarizations refer respectively to electric fields parallel and perpendicular to the plane of incidence on the sample.

Our amorphous and ($z$-cut) crystalline $SiO_2$ samples have a $1$mm thickness and are optically-polished on both sides, provided by Surfacenet GmbH. For amorphous silica, its optical index at $800nm$ is $n_{am} = 1.4533$ \cite{Malitson1965, Arosa2020} while the crystalline quartz is anisotropic, i.e. $n_x = n_y = n_0 = 1.5383$ for the in-plane index and $n_z = n_e = 1.5472$ for the out-of-plane index\cite{Ghosh1999}.

\subsection{Measurement procedure of the probe's polarization rotation}

Under the influence of the intense pump, the probe polarization may change due to pump-induced modifications of the sample’s optical properties. First, its electric field can rotate toward either $P$ or $S$, corresponding to a shift of $\theta$ to $\theta + \Delta\theta$, which can be quantified by so-called Kerr rotation measurements \cite{Yang1993}. Additionally, the linear polarization may evolve toward an elliptical form, which can be characterized through ellipticity measurements \cite{vonVolkmann2009}. However, due to time constraints, only Kerr rotation measurements were performed in this study.\\

\begin{figure}[htbp]
    \centering
    \includegraphics[width=12.5cm]{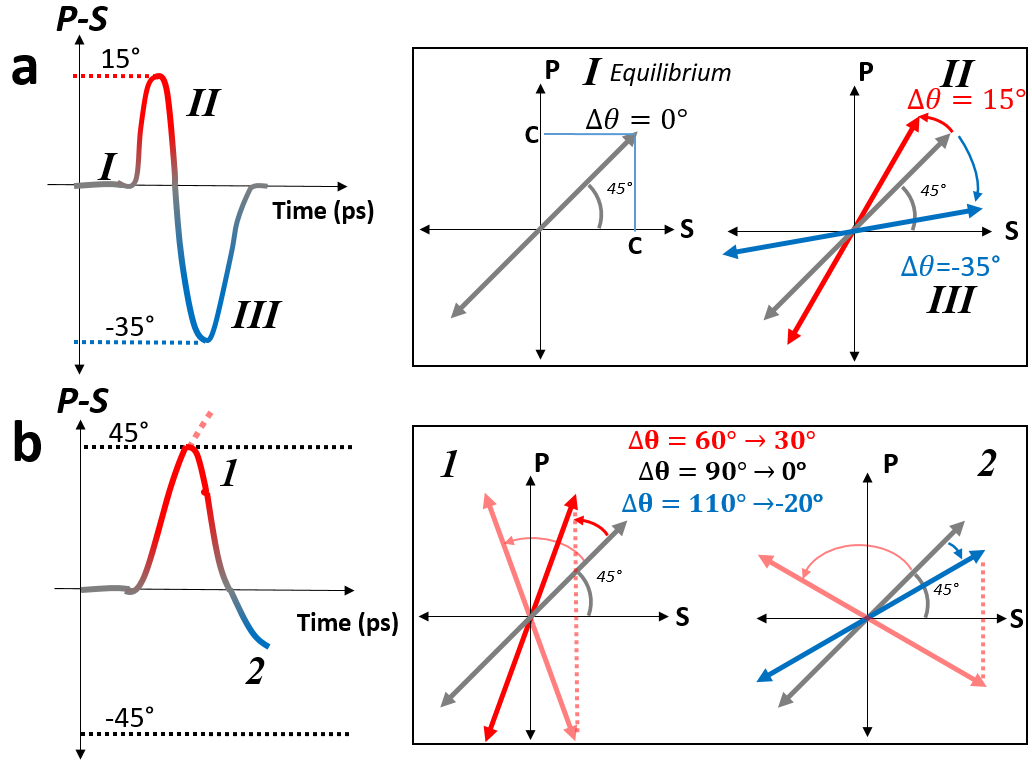}
    \caption{Principle of the Kerr polarisation measurements.}
    \label{fig2}
\end{figure}

To extract the value of $\Delta\theta$, the reflected (or transmitted) probe is analyzed by two photodiodes that receive the $P$ and $S$ components, separated by a Wollaston prism. A lock-in amplifier records the differential signal ($P-S$). A half-waveplate  placed between the sample and the prism ensures that, at equilibrium ($\theta = 45^{\circ}$), the two detectors receive equal intensities, yielding $P-S = 0$. This optical calibration allows for the determination of the sign of $\Delta\theta$ and thus the direction of the polarization rotation (clockwise or counterclockwise), as shown in Figure \ref{fig2}a.  Without this adjustment, detecting the rotation direction would be impossible if the beam were fully aligned with either $P$ or $S$ at equilibrium. 

Denoting by $C$ the equilibrium intensity on one photodiode, we obtain $P = 2C\cos^2(45^{\circ} + \Delta\theta)$, $S = 2C\sin^2(45^{\circ} + \Delta\theta)$, and $S+P = 2C$, implying that the $S-P$ voltage signal can be converted into the angle value $\Delta\theta$ using the relation:

\begin{equation}
    \Delta\theta = \frac{1}{2}\arcsin\left(\frac{S-P}{S+P}\right) 
\end{equation}

However, a geometric limitation arises for $|\Delta\theta| > 45^{\circ}$, where the photodiodes cannot record negative intensities. In that case, due to this geometrical constraint, the real value $\Delta\theta_{real} = 45^{\circ} + \Delta\alpha > 45^{\circ}$ will be measured at a lower value  $\Delta\theta_{mes} = 45^{\circ} - \Delta\alpha < 45^{\circ}$ (and similarly for the negative branch).

For instance, a real $\Delta\theta_{\mathrm{real}} = 60^{\circ}$ is measured as $\Delta\theta_{\mathrm{mes}} = 30^{\circ}$. For $\Delta\theta_{real} > 90^{\circ}$ a negative $\Delta\theta_{mes}$ value will be acquired. Finally, this effect of truncated values induced by the measurement must be taken into account when comparing with the outcome of the simulations in which the real value of $\Delta\theta$ will be calculated.\\

It is worth noting that a potential experimental limitation may arise at high excitation intensities. In our setup, two lock-in amplifiers were used to record the $S-P$ voltage on the reflectivity and transmissivity arms, respectively, while the $S+P$ voltage was not systematically monitored at every time delay. During the initial measurements at $100mJ/cm^2$, we verified that the $S+P$ voltage remained stable for various pump and probe polarizations at two representative time delays — one before time zero and one near it, where the signal amplitude is maximal. Based on these observations, the $S+P$ voltage was assumed to remain constant and was therefore only checked at the beginning of each measurement and occasionally during acquisition. However, at higher excitation intensities, variations of the $S+P$ voltage may occur, possibly due to significant absorption of the probe beam.

\section{Results}

\subsection{Dependence on the relative pump and probe polarization angle $\theta_{ps}$}

\begin{figure}[htbp]
    \includegraphics[width=12.2cm]{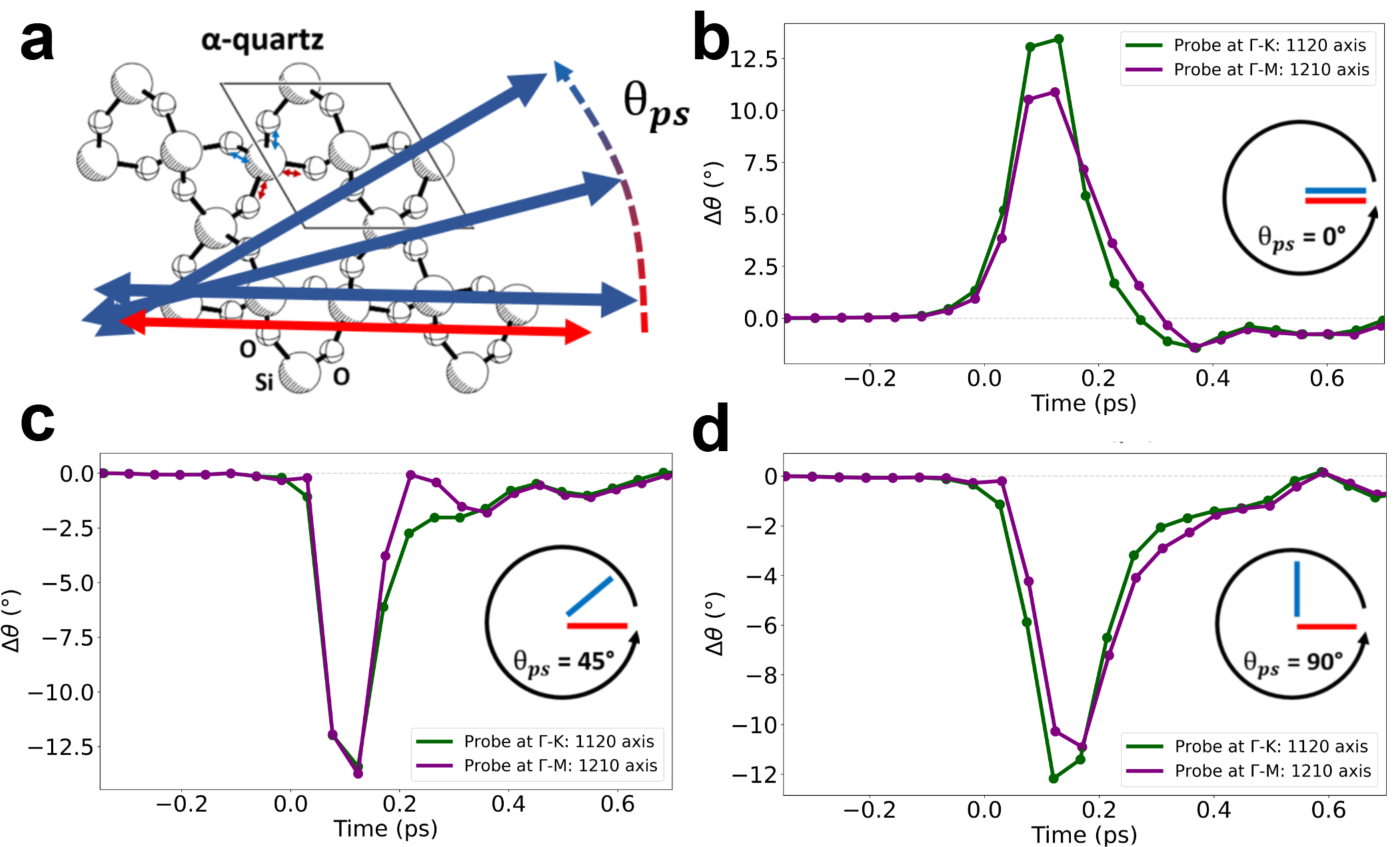}
    \caption{a) Schematic representation of relative angle $\theta_{ps}$ between pump and probe polarisation on the $\alpha$-quartz sample. Here the probe is toward the 1120 axis ($\Gamma-K$). b, c, d) Time-resolved reflectivity measurement on $\alpha$-quartz at $500mJ/cm^2$ of $\Delta\theta$, the photo-induced polarisation angle change, for three values of $\theta_{ps}$ at $0^{\circ}$, $45^{\circ}$ and $90^{\circ}$. The data in green and purple are taken for two different high-symmetry orientations of the probe polarisation on the sample, namely $\Gamma-K$ (1120 axis) and $\Gamma-M$ (1210 axis).}
    \label{fig3}
\end{figure}

The out-of-equilibrium dynamics of quartz are fathomed by varying the pump fluence and/or the relative angle of the polarizations of the probe and pump beams, within the $P$/$S$ plane, denoted as $\theta_{ps}$. As depicted on Figure \ref{fig3}a, the probe polarization is fixed along a high-symmetry direction of the quartz c-plane - either the [1120] or [1210] cristallographic axis - corresponding in the reciprocal space to the $\Gamma-K$ and $\Gamma-M$ directions, as drawn in Figure \ref{fig3}a. The pump polarization is then rotated by an angle $\theta_{ps}$ relative to that of the probe. 

Figure \ref{fig3}b-d shows the time-resolved reflectivity measurements of the probe polarization rotation $\Delta\theta$ for three configurations, $\theta_{ps}=0^{\circ}$, $45^{\circ}$ and $90^{\circ}$, at a pump fluence of $500\,\mathrm{mJ/cm^2}$. Two features are visible: an oscillation from the $3.8$THz coherent optical phonons \cite{Zhang2024, Bosak2012}, and a sharp peak around time zero associated with the optical Kerr effect (OKE), on which we focus. The displayed zero delay is slightly shifted from the true one due to imperfect calibration. However, the time zero is the same for the displayed data.

The OKE signal shows a clear dependence on $\theta_{ps}$: it is positive at $\theta_{ps}=0^{\circ}$ and negative at $45^{\circ}$ and $90^{\circ}$. Rotating the probe between the two high-symmetry directions yields no qualitative change, indicating that the relevant parameter is the relative angle $\theta_{ps}$ rather than the probe orientation with respect to the crystal. Even if hot carriers in $SiO_2$ drift along a preferred crystallographic direction, the probe does not measure this current but the pump-induced perturbation of the dielectric tensor, which is primarily governed by the pump polarization. Moreover, since the electronic gap of quartz lies at the $\Gamma$ point \cite{Chelikowsky1977, Platt2005, Jrgens2019}, the pump does not select any crystallographic direction, and the same applies to the polycrystalline sample. Consequently, the response depends only on $\theta_{ps}$ between pump and probe, motivating our choice to place the probe along $\Gamma$-K in the measurements acquired after this one.

Nevertheless, if the carriers drift along a preferred direction and form a transient plasmon, one may expect this anisotropic charge motion to leave a signature in the temporal response. In our measurements, the signal appears slightly earlier when the probe is aligned along the $\Gamma-K$ direction compared with the $\Gamma-M$ direction. This timing asymmetry suggests that the photoexcited carriers—and therefore the transient DC field they generate—are more efficiently driven or stabilized along the $\Gamma-K$ axis, consistent with a drift or effective-mass anisotropy in that direction. However, this observation could also be induced by a setting of the $\theta_{ps}$ angle which is not exactly the same between the two measurements.

\subsection{Sign symmetry of the Kerr effect response with respect to $\theta_{ps}$}

\begin{figure}[htbp]
    \centering
    \includegraphics[width=13.5cm]{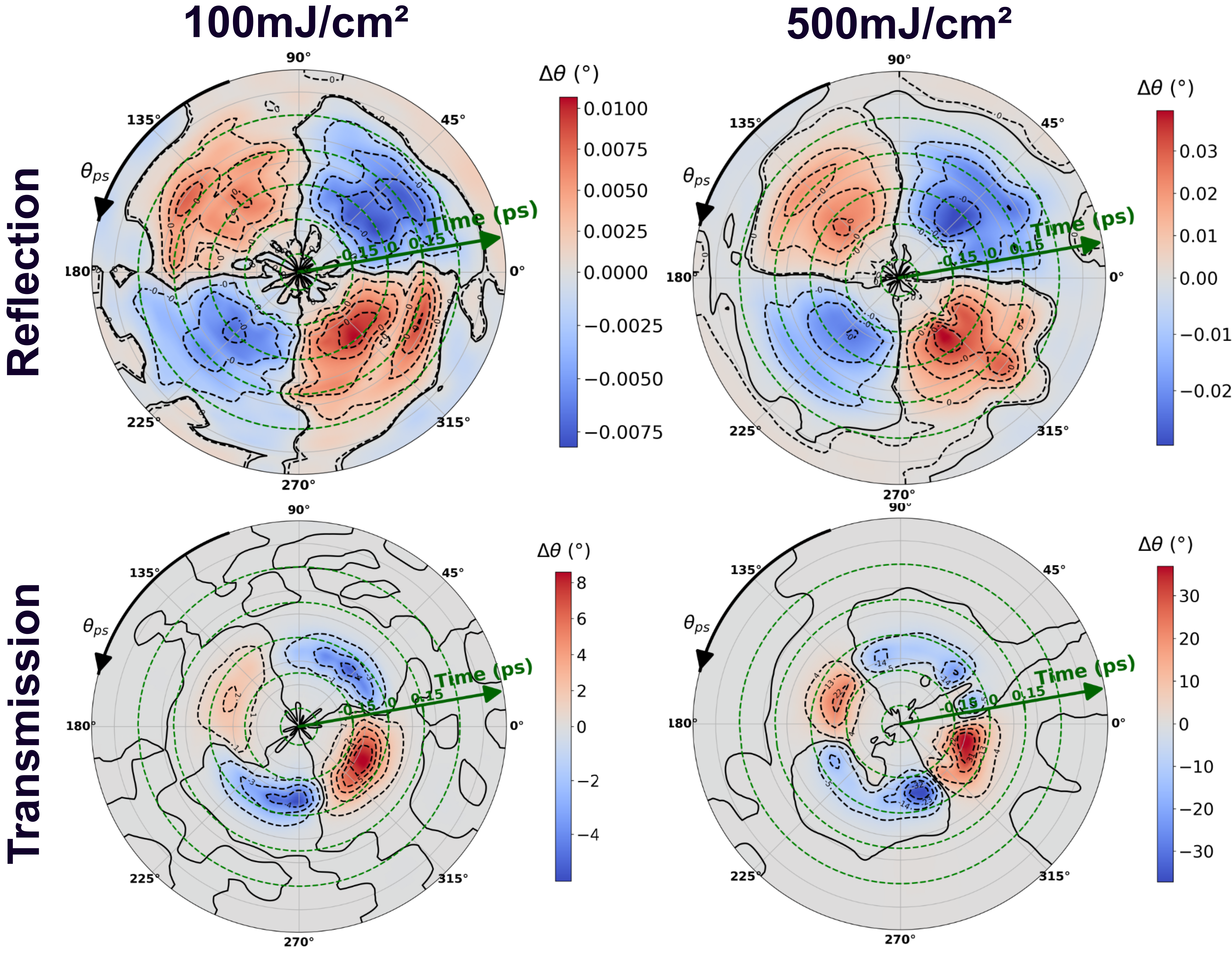}
    \caption{Amorphous $SiO_2$ : Symmetries of the time-resolved signals in reflection and transmission for two pump fluences at $100$ and $500mJ/cm^2$. All polar graphs have time delay for radial axis (in green), relative angle $\theta_{ps}$, for the azimuthalaxis and $\Delta\theta$ values for the colorbar. These graphs are referenced as $\Delta\theta(t,\theta_{ps})$ graphs. The registered polarisation angle change $\Delta\theta$ is scaled with blue-grey-red colors for negative-zero-positive values.}
    \label{fig4}
\end{figure}

\begin{figure}[htbp]
    \centering
    \includegraphics[width=12cm]{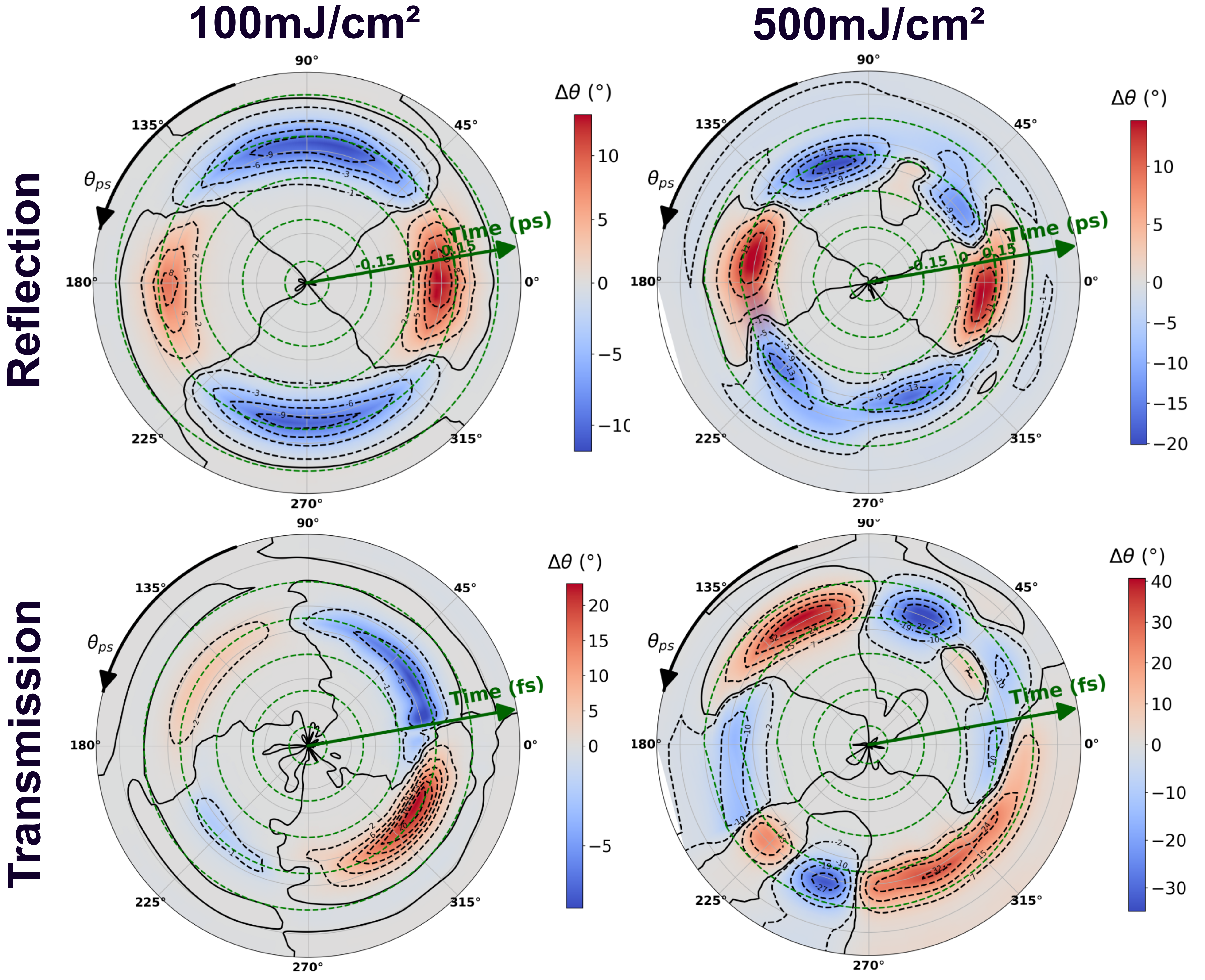}
    \caption{$\alpha$-$SiO_2$ : Symmetries of the time-resolved signals in reflection and transmission for two pump fluences at $100$ and $500mJ/cm^2$. All polar graphs have time delay for radial axis (in green), relative angle $\theta_{ps}$, for the azimuthalaxis and $\Delta\theta$ values for the colorbar. These graphs are referenced as $\Delta\theta(t,\theta_{ps})$ graphs. The registered polarisation angle change $\Delta\theta$ is scaled with blue-grey-red colors for negative-zero-positive values.}
    \label{fig5}
\end{figure}

The sign of the OKE signal with respect to $\theta_{ps}$ has a particular symmetry and exhibits clear dependencies on both pump fluence and the detection geometry, i.e. reflection or transmission. Figures \ref{fig4} and \ref{fig5} display these measured symmetries for the amorphous silica and $\alpha$-quartz samples respectively. 

We use polar plots referenced as $\Delta\theta(t,\theta_{ps})$ graphs, where the radial axis represents the pump-probe time delay, the azimuthal angle corresponds to $\theta_{ps}$, and the color scale encodes the amplitude of the polarization rotation. We performed these measurements for 27 values of $\theta_{ps}$ (every $10^{\circ}$ between first $-100^{\circ}(260^{\circ})$ and $+100^{\circ}$, and then every $30^{\circ}$ elsewhere).

\paragraph{Low fluence observations ($100mJ/cm^2$):}

At low fluence ($100mJ/cm^2$), the sign of the OKE response exhibits a fourfold symmetry with respect to the angle $\theta_{ps}$. However, it differs if the measure is in reflection or transmission:

\begin{itemize}
    \item In transmission, a clear fourfold symmetry is observed with sign inversions of $\Delta\theta$ occurring at $\theta_{ps}^{inv,T} = 0^{\circ}$, $90^{\circ}$, $180^{\circ}$ and $270^{\circ}$, and with extrema appearing for pump and probe polarizations at $\pm 45^{\circ}, \pm 135^{\circ}$.
    \item In contrast, the symmetry in reflection is distinct between amorphous silica and the $\alpha$-quartz. For fused silica, its symmetry is the same as in transmission, while for the $\alpha$-quartz case the OKE response $\Delta\theta$ is positive for collinear pump and probe polarizations ($\theta_{ps} = 0^{\circ}$ and $180^{\circ}$), changes sign near $\theta_{ps} \approx \pm 35^{\circ}$ and reaches minima for the cross-polarized configuration at $\theta_{ps} \approx \pm 90^{\circ}$. Moreover, the amplitude of $\Delta\theta$ is totally different between the two samples: of about $10^{\circ}$ for quartz while the amplitude is 1000 times lower for amorphous silica. 

    Another observations in reflection must be noticed: in amorphous silica, a second peak is observed $\sim200$fs after the first one (this delay is independent on the fluence and the angle $\theta_{ps}$). This was totally unexpected will be discussed at the end of this article. Furthermore, for the $\alpha$-quartz, the intensity is increasing slightly slower with time delay than its latter drop. This might be linked to a retarded effect such as due an effect induced by conduction electrons as it takes about $50$-$100$fs to photo-excite all of them \cite{Jrgens2019}.
\end{itemize} 

A key objective of this article is to explain why such symmetries are observed in transmission and reflection, and try to explain why such a big difference between amorphous and crystalline $SiO_2$ occurs in reflectivity. 

Finally, the drop in amplitude for $\theta_{ps}$ between $90^{\circ}$ and $270^{\circ}$ is attributed to a gradual drift of the laser pointing during the acquisition, which is more pronounced for these last measurements and might has been slightly enforced by pump-probe Light-Induced Transient Grating (LITG) \cite{Yue2009}.

\paragraph{Higher fluence observations ($500mJ/cm^2$):}

At higher fluence ($500mJ/cm^2$), the symmetry becomes distorted at several $\theta_{ps}$ angles. Some of these distortions likely occur when $|\Delta\theta| > 45^{\circ}$, causing the measured value to deviate from the actual one. For instance,  in figure \ref{fig5}, in transmission at $\theta_{ps} = 45^{\circ}$ or reflection for $\theta_{ps} \approx 90^{\circ}$, a sign reversal is observed, implying a real $|\Delta\theta| > 90^{\circ}$. This effect is discussed further in Section 4.1.2.

Moreover, we note that the fluence induces a change in the transmitted signal at parallel polarization ($\theta_{ps} = 0^{\circ}, 180^{\circ}$). At low fluence, a negative peak is followed by a positive one, whereas at high fluence only a single positive peak is clearly observed.

\paragraph{Long lifetime of the signal dynamics:}

If we focus on the lifetime of the signal dynamics $\Delta\theta(t)$, i.e. the timescale over which the Kerr peak remains observable, we find that it extends over a rather long duration compared to theory, on the order of 150–250 fs, and increases with fluence. In contrast, the Kerr effect is typically regarded as instantaneous with respect to the pump–material interaction, meaning it should follow the same temporal shape as the pump pulse convoluted with the probe. This would yield a theoretical lifetime of only about 70–100 fs, as illustrated in Figure \ref{fig6}a (green curve).

\section{Discussion}

\subsection{Modelisation of the transmission with phase-retardation approach}

\subsubsection{Analytical formalism} 

We model the transmission symmetry using a simple phase-retardation model. The Kerr induced birefringence $\Delta n$ alters the probe polarization as it travels through the sample’s excited region. This birefringence induces a phase shift $\Delta\phi$ given by: 

\begin{equation}
    \Delta\phi = \frac{2\pi d}{\lambda} \Delta n = \frac{2\pi d}{\lambda} n_2\frac{F}{\delta t}
    \label{eq:delta_phi}
\end{equation}

with $\lambda = 800$nm the probe wavelength, $n_2 = 3.4\times10^{-20}\mathrm{m^2/W}$ the Kerr coefficient \cite{Santran2004}. $F$ and $\delta t=50$fs are the pump fluence and pulse duration, while the ratio $F/\delta t$ is the pump intensity.\\

The length $d$ denotes the effective overlap between the pump-excited region \cite{Mao2004, bilde:tel-02169576} and the probe path. As suggested by Eq.~\ref{eq:delta_phi}, we treat $F$ as uniform along this path, effectively assuming a depth-independent excitation. Under this approximation, $d$ behaves as an effective interaction length—the distance a probe would travel inside a uniformly excited medium.

In practice, the excitation is not uniform: the pump has a finite transverse size, and its intensity decays significantly during propagation \cite{bilde:tel-02169576}. Supplementary Information~1 introduces a spatial-overlap model accounting for this inhomogeneity, yielding $d \approx 70\mu m$ for our conditions. For pump fluences of $100$ and $500,\mathrm{mJ/cm^2}$, we then estimate $\Delta\phi \simeq 0.119\pi$ and $0.595\pi$, respectively.

However, the excited-depth profile used here is taken from simulations in $\alpha$-quartz for a fluence around $700,\mathrm{mJ/cm^2}$ \cite{bilde:tel-02169576}. Since this profile depends on fluence, a lower fluences produce deeper, less depleted excitation. Thus, the effective length $d$ should increase at lower $F$, naturally enhancing the observed Kerr signal.

Assuming the pump is polarized along the arbitrary $x$-axis, and the probe at $\theta_p = -\theta_{ps}$, its electric field $E_p$ acquires a modified phase in the $x$-direction, denoted as $\hat{x}$, equivalent to a polarization rotation by an angle $\Delta\theta$, such that (with $E_0(t) = E_0 e^{j\omega t}$):

\begin{equation}
    E_p (t) = E_0(t) \cos(\theta_p) e^{j\Delta\phi} \hat{x} + E_0(t) \sin(\theta_p) \hat{y}
    = E_0(t) \cos(\theta_p + \Delta\theta) \hat{x} + E_0(t) \sin(\theta_p + \Delta\theta) \hat{y} 
\end{equation}

Leading to $\tan(\theta_p + \Delta\theta) = tan(\theta_p)/ e^{j\Delta\phi}$, implying that $\Delta\theta$ is complex, and thus denoted as $\overline{\Delta\theta}$ in the following. To isolate $\overline{\Delta\theta}$, we use the trigonometric identity:

\begin{equation}
    \tan(\overline{\Delta\theta}) = \tan(\theta_p + \overline{\Delta\theta} - \theta_p) =  \frac{\tan(\theta_p + \overline{\Delta\theta}) -\tan(\theta_p)}{1 + \tan(\theta_p + \overline{\Delta\theta})\tan(\theta_p)} 
\end{equation}

We finally consider that the measured rotation $\Delta\theta$ is linked to $\mathcal{R}e(\Delta\theta)$ for its amplitude and $\mathcal{I}m(\Delta\theta)$ for its sign, such that 

\begin{equation}
    \Delta\theta = \mathcal{R}e\left[\arctan(\tan(\overline{\Delta\theta}))\right] \times \mathcal{S}ign\left(\mathcal{I}m\left[\arctan(\tan(\overline{\Delta\theta}))\right]\right)
\end{equation}

\begin{figure}[htbp]
    \centering
    \includegraphics[width=12cm]{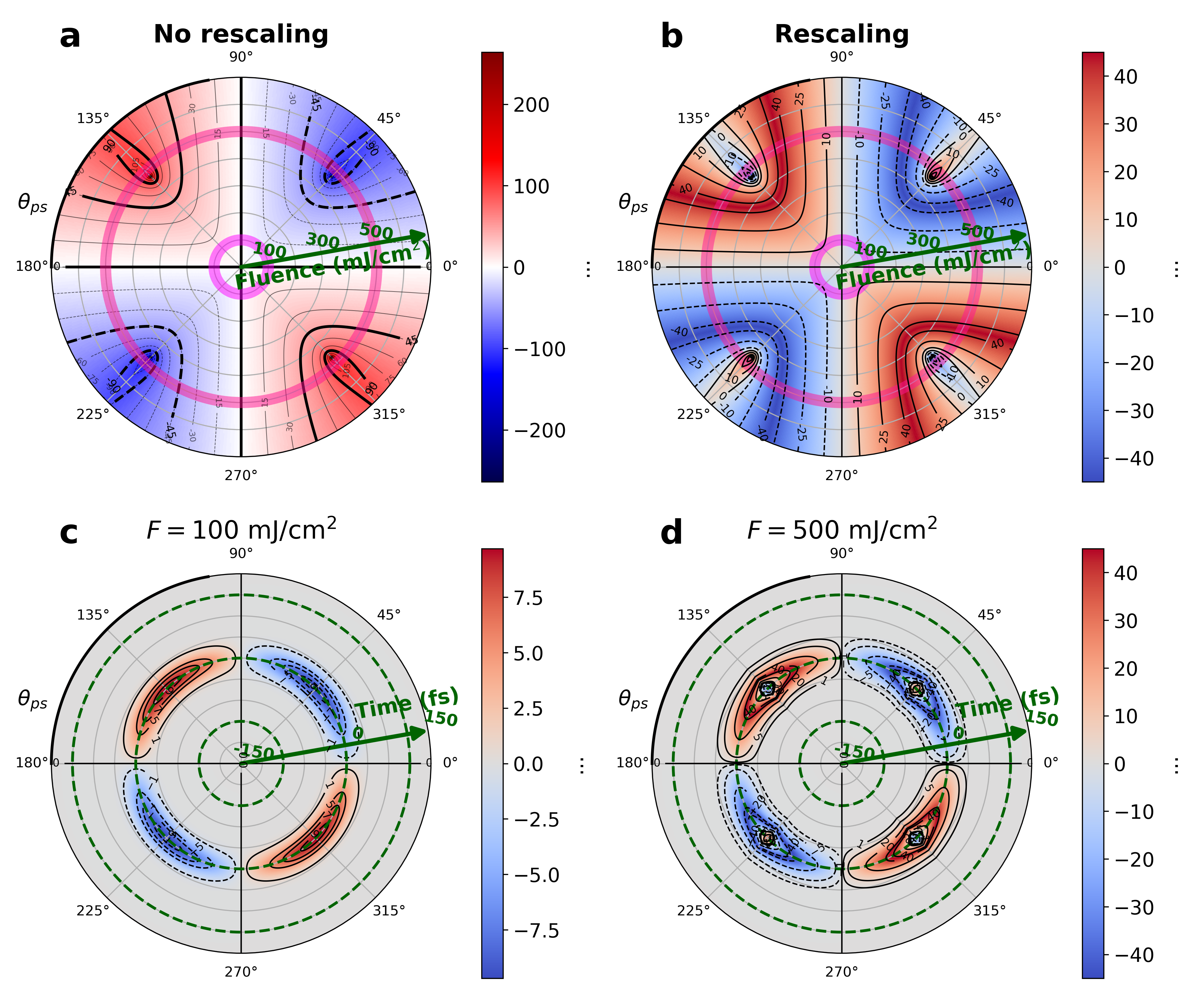}
    \caption{Simulation of $\Delta\theta$’s symmetry in transmission with induced birefringence and  phase delay formalism. a) Simulation of the $\Delta\theta(F,\theta_{ps})$ graph displaying the change of polarisation $\Delta\theta$ with respect to the pump fluence $F$ (radial axis) and the polarisation configuration angle $\theta_{ps}$ (azimuthalaxis). The values of $\Delta\theta = \pm 45^{\circ}$ and $\pm 90^{\circ}$ are enlighted with black countours and the $100$ and $500mJ/cm^2$ fluences are indicated by the purple concentric circles. b) Rescaled $\Delta\theta(F,\theta_{ps})$ figure for $\Delta\theta$ values adjusted according to the measurement acquisition. c and d) Simulation of the $\Delta\theta(t,\theta_{ps})$ graphs at $100$ and $500mJ/cm^2$ pump fluences respectively.}
    \label{fig6}
\end{figure}

The simulated value of $\Delta\theta$ is then displayed on Figure \ref{fig6}a on a polar graph with fluence for the radial axis. On this one we observe that, around $\theta_{ps} = \pm 45^{\circ}, \pm 135^{\circ}$, for a fluence close to $350mJ/cm^2$($450mJ/cm^2$) we predict $|\Delta\theta| > 45^{\circ}$($90^{\circ}$), implying a rescaling of the measured data compared to the real one, which is presented in Figure \ref{fig6}b. From this, we reconstruct the $\Delta\theta(t,\theta_{ps})$ maps at $100$ and $500~\mathrm{mJ/cm^2}$ by setting the non-rescaled $\Delta\theta$ value as the peak of a Gaussian profile with a $FWHM = 50$fs, corresponding to the pump pulse duration and therefore to an instantaneous OKE response. This Gaussian profile is then rescaled, yielding the simulated $\Delta\theta(t,\theta_{ps})$ maps shown in Figure~\ref{fig6}c–d.

The low-fluence symmetry of the OKE signal is accurately reproduced, both in sign and in amplitude. At higher excitation levels, the overall profile is still reproduced; however, the agreement becomes less satisfactory, with two main discrepancies emerging.

The first concerns an expected sign reversal around $\theta_{ps} \approx \pm 45^{\circ}$ and $\pm 135^{\circ}$, which arises when the real amplitude reaches $|\Delta\theta| > 90^{\circ}$. In our measurements, only the configurations at $\theta_{ps} = 45^{\circ}$ and $135^{\circ}$ were recorded. According to the model, the signal should evolve from positive to negative and then back to positive, whereas experimentally only a negative-to-positive transition is observed. This behavior is most likely a consequence of the limited temporal resolution of our setup ($\sim 50$fs), but might also be linked to the $10^{\circ}$ tilt of the incoming probe beam on the sample. The second discrepancy appears for the special case of parallel pump–probe polarizations ($\theta_{ps} = 0^{\circ}$ and $180^{\circ}$). Within our model, no probe polarization change is expected in this configuration. Nevertheless, a distinct signal is observed at both fluences (Figure \ref{fig5}, bottom), consisting of two opposite peaks — first positive, then negative — with the positive component becoming dominant as the excitation increases. \\

A likely primary cause of the high-intensity discrepancies is an experimental artifact, as the $P+S$ voltage may vary when the sample strongly absorbs the probe beam. Other physical effects could also contribute. Nonlinear photoexcitation may become significant at high fluence, enhancing pump absorption and causing its intensity to decay during propagation. Additional high-intensity effects, such as self-phase modulation, can introduce temporal modulations or double-peak structures not captured by our simple model. Furthermore, the behavior may reflect higher-order nonlinearities, temporal beam splitting induced by self-phase modulation \cite{Diddams1998} or a dynamic modification of the pump polarization within the OKE process. Indeed, transient variations in pump polarization angle or ellipticity have been observed \cite{Gertsvolf2010}, effectively altering $\theta_{ps}$ during the delay, which could explain this non-zero signal at collinear configuration.

\subsubsection{Retarded Kerr effect}

One striking feature of our measurements is the relatively long lifespan of the signal dynamics $\Delta\theta(t)$, i.e. the timescale over which the Kerr peak remains observable. While theory would suggest durations of only $\sim 70$–$100$fs (from the convolution of the pump and probe beams), we measure signals extending up to $150$–$250$fs (see Figure \ref{fig5}). Explaining this discrepancy is the one of the aim of this section.

We focus on the case $\theta_{ps}=45^{\circ}$. Here, $\Delta\theta(t)$ initially exhibits a positive sign, which rapidly switches to negative within a few tenths of a femtosecond before relaxing back to zero. This can be explained by the dynamics of the OKE response and its truncation due to the measurement acquisition. Indeed, the real value of $\Delta\theta$ encounters a rapid increase and drop in amplitude, but the truncation induces a change of its sign while the OKE response is of maximal intensity. Then, when the signal relaxes, we observe a sign change of the registered signal.\\

To reproduce such a dynamics, we model $\Delta\theta(t)$ as a time-delayed response of the optical Kerr effect (OKE) convoluted with a Gaussian probe of FWHM = $50$fs:

\begin{equation}
    \Delta\theta(t) = OKE(t) \circledast Gauss_{probe}(t)
\end{equation}

First, we assume an instantaneous OKE, i.e. with the same Gaussian shape as the pump pulse:

\begin{equation}
    OKE_{inst}(t) = A e^{-\frac{t^2}{2\sigma^2}} \quad\quad\quad \text{with} \quad\sigma = \frac{FWHM}{2\sqrt{2\log(2)}}
\end{equation}

\begin{figure}[htbp]
    \centering
    \includegraphics[width=14cm]{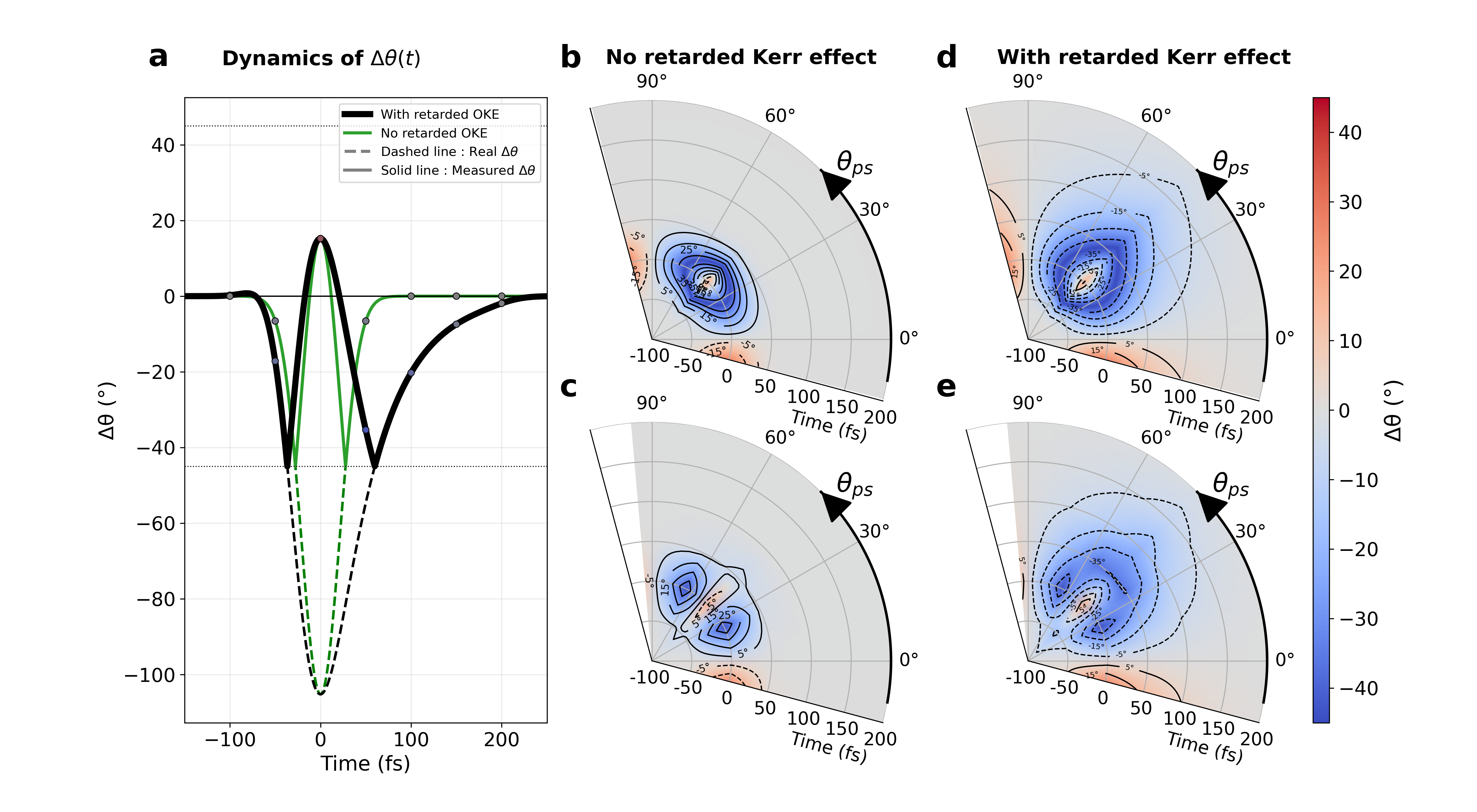}
    \caption{Retarded OKE to explain the observed signal in transmission at $\theta_{ps} = 45^{\circ}$. a) Dynamics of the signal of $\Delta\theta$ with and without considering a retarded OKE. b) Simulation of the $\Delta\theta(t,\theta_{ps})$ signal without retarded OKE. c) Same simulation but with a discretization in angles (every $15^{\circ}$) and time delay (50fs) reproducing the data acquisition. d,e) Same but with retarded OKE of $\tau = 25fs$.}
    \label{fig7}
\end{figure}

The corresponding signal is shown in Figure \ref{fig7}a, with green dashed and solid lines representing the real and rescaled $\Delta\theta(t)$, respectively. Figure \ref{fig7}b presents the $\Delta\theta(t,\theta_{ps})$ map, while Figure \ref{fig7}c displays a coarsely discretized version in both $\theta_{ps}$ and time delay, sampled every $50$fs, matching the experimental acquisition. This temporal resolution is insufficient to reproduce the experimentally observed signal, as no sign change clearly appears at $\theta_{ps} = 45^{\circ}$.\\

To address this limitation, we consider a retarded OKE response \cite{43}. The OKE corresponds to a pump-induced polarization of the medium, which may be delayed depending on the involved microscopic mechanisms. The standard Kerr effect is linked to the ultrafast electronic cloud deformations on a timescale $h/\Delta_{gap} \approx 0.5$–$5$ fs \cite{44}, which is too brief to explain our signal. Another probable origin is the formation of a persistent plasma that polarizes the medium. In $SiO_2$, the ultrafast formation and relaxation of a plasmon within a hundred femtoseconds has been proposed to explain certain experimental observations at higher fluences \cite{Mller2020}, and could be related to our observed delayed response. Nevertheless, a Raman origin of the OKE signal cannot be excluded, since in our intensity regime the pump beam may displace ions from their equilibrium positions. These ions would then require a finite time to relax back to equilibrium — typically on the order of 40–80 fs in gases and liquids \cite{41,45,46} — suggesting a phononic nature of the retarded OKE response.

We thus test a model as the convolution of the instantaneous response with a causal exponential decay $\tau = 25fs$, weighted by the Heaviside function $H(t)$ to ensure causality. This formulation has an analytical solution with use of the error function:

\begin{equation}
    OKE_{ret}(t) = A\left(\frac{1}{\sqrt{2\pi}\sigma}e^{-\frac{t^2}{2\sigma^2}}\right) \circledast \left(\frac{H(t)}{\tau}e^{-\frac{t}{\tau}}\right) = \frac{A}{2\tau}e^{\left(\frac{\sigma^2}{2\tau^2} - \frac{t}{\tau}\right)} \times erf\left(\frac{\sigma}{\tau\sqrt{2}} - \frac{t}{\sigma\sqrt{2}}\right)
    \label{eq:retarded_kerr}
\end{equation}

This formulation introduces a relaxing tail to the OKE response (see Figure \ref{fig7}a), consistent with both the long lifespans and the observed sign alternation in $\Delta\theta(t)$ at $\theta_{ps} = 45^{\circ}$ (see Figure \ref{fig7}e) for our measurement resolution. In the following, all the $\Delta\theta(t,\theta_{ps})$ graphs will be described with this retarded OKE dynamics.

\subsection{Modelisation of the reflection with an anisotropic tensorial model}

\subsubsection{Anisotropic tensorial model}

To model the angular dependence of the pump–probe optical Kerr effect, we develop a tensorial description of the pump-induced permittivity in amorphous silica and in $\alpha$-quartz. A standard approach for anisotropic media is to start from the optical impermeability \cite{shen, nye1985physical, Steglich2024},

\begin{equation}
    B_{ij}=\bigl(1/n^{2}\bigr)_{ij}
\end{equation}

which acquires both linear (Pockels) and quadratic (Kerr) electro-optic corrections under an applied electric field $E_k$,

\begin{equation}
    \left(\frac{1}{n_{ij}^{2}}\right)_{E}
    =
    \left(\frac{1}{n_{ij}^{2}}\right)_{0}
    + r_{ijk}E_{k}
    + g_{ijkl}E_{k}E_{l}
\end{equation}

Amorphous fused silica is centrosymmetric: its Pockels tensor vanishes ($r_{ijk}=0$), so the pump-induced birefringence arises solely from the isotropic Kerr tensor $g_{ijkl}$. In contrast, $\alpha$-quartz (point group 32) exhibits both Pockels and Kerr contributions \cite{nye1985physical, Glytsis2002LectureNotes}, and is already anisotropic in the unpumped state. In the following, we focus on fused silica, which is central to the interpretation of our measurements. Complete modelization for both materials are provided in Supplementary Information 2.

% =====================================================
\paragraph{Pump-induced permittivity of amorphous silica}
% =====================================================

In the experiment, the pump polarization is rotated in the $(x,y)$ plane by an angle $\theta_{ps}$ relative to the probe. The pump field is therefore written as

\begin{equation}
    \mathbf E = E(\cos\theta_{ps},\sin\theta_{ps},0)
\end{equation}

In an isotropic medium, the fourth-rank tensor $g_{ijkl}$ reduces to two independent coefficients, $s_{11}$ and $s_{12}$. The coefficient $s_{11}$ is directly related to the usual scalar Kerr index $n_{2}$ (see Supplementary Information 2), and $s_{12}$ is manually tuned in order to have a good agreement with the measurements. We then have $s_{11} = -4.2\times 10^{-23}m/V$ and $s_{12} \approx 1.5\times 10^{-22}m/V$. Having $|s_{11}| < |s_{12}|$ is suprising as we do not expect to have a transverse effect stronger than the collinear one. For an in-plane pump field, the Kerr-induced changes of optical impermeability are

\begin{equation}
    \begin{aligned}
    \Delta B_{xx}
    &= E^{2}\bigl(s_{11}\cos^{2}\theta_{ps}
    + s_{12}\sin^{2}\theta_{ps}\bigr)\\[2pt]
    \Delta B_{yy}
    &= E^{2}\bigl(s_{12}\cos^{2}\theta_{ps}
    + s_{11}\sin^{2}\theta_{ps}\bigr)
    \end{aligned}
    \qquad
    \begin{aligned}
    \Delta B_{xy}
    &= \tfrac14(s_{11}-s_{12})E^{2}\sin(2\theta_{ps})\\[2pt]
    \Delta B_{zz} &= s_{12}E^{2}\qquad
    \Delta B_{xz}=\Delta B_{yz}=0
    \end{aligned}
\end{equation}

\begin{minipage}[c]{0.48\linewidth}
        Inverting the impermeability matrix tensor and expanding all terms to second order in $E$ yields the relative permittivity tensor in the $(x,y,z)$ frame:
\end{minipage}
\hfill
    % --- Second image ---
\begin{minipage}[c]{0.48\linewidth}
    \begin{equation}
        \boldsymbol\varepsilon^{(am)}_{(x,y,z)} = \varepsilon_0 
        \begin{pmatrix} 
        \varepsilon_{r,xx} & \varepsilon_{r,xy} & 0\\ 
        \varepsilon_{r,xy} & \varepsilon_{r,yy} & 0\\ 
        0 & 0 & \varepsilon_{r,zz} 
        \end{pmatrix} 
    \end{equation}
\end{minipage}

with the components

\begin{equation}
    \begin{aligned}
    \varepsilon_{r,xx}
    &= n_{am}^{2}
    - n_{am}^{4}E^{2}
    \bigl(s_{11}\cos^{2}\theta_{ps}
    + s_{12}\sin^{2}\theta_{ps}\bigr)\\[2pt]
    \varepsilon_{r,yy}
    &= n_{am}^{2}
    - n_{am}^{4}E^{2}
    \bigl(s_{12}\cos^{2}\theta_{ps}
    + s_{11}\sin^{2}\theta_{ps}\bigr)
    \end{aligned}
    \qquad
    \begin{aligned}
    \varepsilon_{r,xy}
    &= -\tfrac14 n_{am}^{4}(s_{11}-s_{12})E^{2}\sin(2\theta_{ps})\\[2pt]
    \varepsilon_{r,zz} &= n_{am}^{2}-n_{am}^{4}s_{12}E^{2}
    \end{aligned}
\end{equation}

In the probe basis $(p,s,z')$, where the $P$-polarized probe lies along $x$ and is tilted by the refracted angle $\beta$ in the $(y,z)$ plane, the relevant in-plane components are

\begin{equation}
    \varepsilon_{pp}=\varepsilon_{xx},\qquad
    \varepsilon_{ps}=\varepsilon_{xy}\cos\beta,\qquad
    \varepsilon_{ss}=\varepsilon_{yy}\cos^{2}\beta+\varepsilon_{zz}\sin^{2}\beta
\end{equation}

% =====================================================
\paragraph{Effective refractive index and refracted angle}
% =====================================================

The pump-induced anisotropy modifies the propagation of the probe inside the medium. The effective index $n_{\mathrm{eff}}$ and refracted angle $\beta$ depend on $E$ and $\theta_{ps}$ through the tensor $\varepsilon_{(p,s,z')}$. Both quantities are obtained using a self-consistent fixed-point procedure detailed in Supplementary Information~3. In practice, the variations remain small ($\sim 10^{-3}$ for $n_{\mathrm{eff}}$ and $\sim 0.1^\circ$ for $\beta$).

% =====================================================
\paragraph{Generalized Fresnel coefficients and polarization rotation}
% =====================================================

A $P$-polarized incident probe is reflected into both $P$ and $S$ components, with reflection coefficients $r_{pp}$ and $r_{ps}$. These arise from the in-plane permittivity tensor

\begin{equation} 
    \boldsymbol\varepsilon_{(p,s)} = 
    \begin{pmatrix} 
        \varepsilon_{pp} & \varepsilon_{ps}\\ 
        \varepsilon_{ps} & \varepsilon_{ss} 
    \end{pmatrix} 
\end{equation}

which is real and symmetric. Diagonalizing it yields two eigenvalues $\varepsilon_{1,2}$ associated with orthogonal optical axes rotated by

\begin{equation}
    \phi
    = \frac12\arctan\left(
    \frac{2\varepsilon_{ps}}{\varepsilon_{pp}-\varepsilon_{ss}}
    \right)
\end{equation}

In this principal-axis frame, the probe reflects through two independent effective isotropic channels, each characterized by a Fresnel coefficient $r_i$ ($i=1,2$) computed using the effective indices $n_{\mathrm{eff},i}$ and refracted angles $\beta_i$ obtained for pump angles $\theta_{ps}=0^\circ$ and $90^\circ$.

Transforming back to the laboratory $(p,s)$ basis yields the measurable coefficients

\begin{equation}
    r_{pp}=r_1\cos^{2}\phi+r_2\sin^{2}\phi
    \qquad
    r_{ps}=\tfrac{1}{2}(r_2-r_1)\sin 2\phi
\end{equation}

Finally, the Kerr-induced rotation of the reflected probe polarization is given by the relation \cite{Yang1993, Heim1996}:

\begin{equation}
    \Delta\theta
    = \Re\left[
    \arctan\left(\frac{r_{ps}}{r_{pp}}\right)
    \right]
\end{equation}

Finally, the same analysis can be performed for the transmissivity (see Supplementary Information 4). While the correct symmetry is recovered, the predicted signal amplitude is about two orders of magnitude smaller than the experimental one, which further supports the validity of our interpretation based on phase retardation.

\subsubsection{Results for amorphous silica}

\begin{figure}[htbp]
    \centering
    \includegraphics[width=13cm]{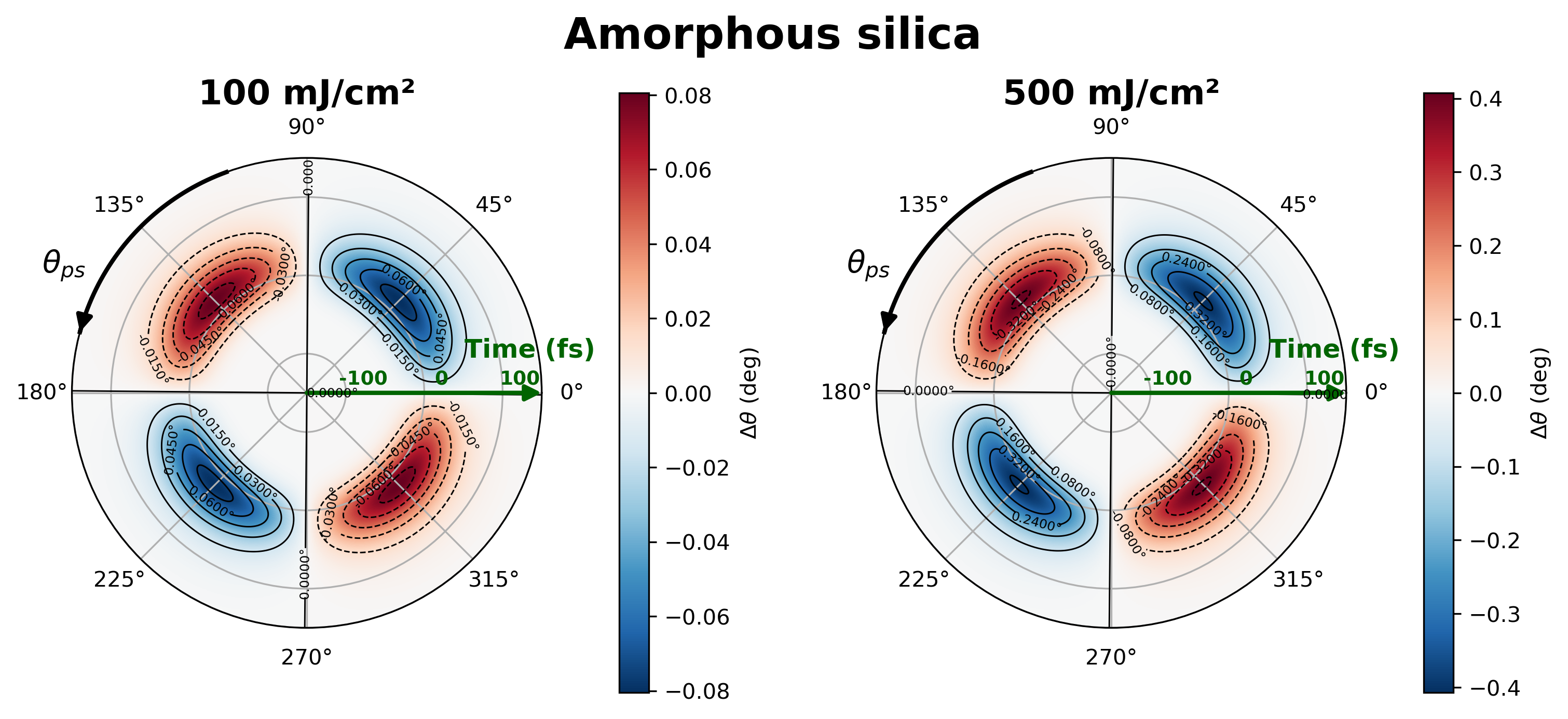}
    \caption{Simulation of the time-resolved $\Delta\theta(t, \theta_{ps})$ maps in amorphous $SiO_2$ with a tensorial permittivity model.}
    \label{fig8}
\end{figure}

In Figure \ref{fig8}, we present the simulated time-resolved scans for amorphous $SiO_2$ at the two fluences considered. A good agreement is obtained with the experimental measurements shown in Figure \ref{fig4}. In particular, the model correctly reproduces the symmetry with respect to $\theta_{ps}$ and the overall amplitude of $\Delta\theta$. However, the secondary peak (occurring approximately 200 fs after the main one) is not reproduced. We propose two hypotheses that may explain its physical origin:

\begin{itemize}
    \item First, once a small anisotropy is established by the OKE, the Kerr response may change because the probe then interacts with an effectively anisotropic Kerr tensor. In addition, the material may acquire a weak Pockels sensitivity. Since the Pockels effect is intrinsically stronger than Kerr, even a tiny pump-induced anisotropy combined with a residual pump field could render this contribution detectable.
    \item A second hypothesis is that, after roughly $100$~fs, the carrier plasma may begin to absorb the probe \cite{Jrgens2019}. 
    Such (non-acquired) absorption, on the order of $0.1$--$1\%$, could differ slightly between $P$ and $S$ polarizations in a transient birefringent medium, thereby inducing a small rotation of the probe polarization. 
\end{itemize}

\subsubsection[Additional piezo-photoelastic effect for alpha-quartz]{Additional piezo-photoelastic effect for $\alpha$-quartz}

Unfortunately, Pockels and Kerr effects are not able to reproduce the $\theta_{ps}$ symmetry observed experimentally in crystalline $SiO_2$. As presented in Supplementary information 4, neither the dominant Pockels effect nor the Kerr contribution yields the correct symmetry or the correct amplitude of the polarization rotation, which is one order of magnitude lower. This indicates that our present description is not sufficient to account for the measurements performed in $\alpha$-quartz.\\

An additional hypothesis is to consider a piezoelectric response, which is known to be significant in quartz. In this scenario, the polarization of the conduction electrons generates a transient, retarded DC electric field that subsequently induces a piezoelectric deformation. As previously mentioned, a long-lived plasmon capable of producing such a polarization has indeed been observed in this material \cite{Mller2020}. The resulting piezoelectric strain then modifies the optical indices through the photoelastic effect. This hypothesis is further supported by the slight increase of the $\Delta\theta$ signal at positive delays, which may correspond to the formation time of this plasmon.

We model this mechanism as described in Supplementary Information~5. We assume that the plasma of conduction electrons creates, in the crystal frame $(x,y,z)$, a DC electric field of magnitude $E_0^{\mathrm{DC}}$ proportional to the photoexcitation rate:

\begin{equation}
    \mathbf{E}^{\mathrm{DC}} = E_0^{\mathrm{DC}} (E_x^{\mathrm{DC}}, E_y^{\mathrm{DC}}, E_z^{\mathrm{DC}})
\end{equation}

This DC field acts on the piezoelectric tensor and produces a strain field $\boldsymbol{\eta}$ \cite{Carr1967, Tiersten1976, AuldBook}. The strain then modifies the optical indicatrix through the photoelastic tensor \cite{NarasimhamurtyBook, Donadio2003}. In Voigt notation, these relations read:

\begin{align}
    \eta_j = \sum_{k=1}^{3} d_{kj}\,E_k^{\mathrm{DC}}
    \qquad\qquad \Delta B_m
    =
    \sum_{n=1}^{6} p_{mn} \,\eta_n
    \qquad (j,m=1,\dots,6)
    \label{eq:piezo-voigt}
\end{align}

where $B_m$ is the impermeability tensor (the inverse of the permittivity tensor) expressed in Voigt notation. Adding this piezoelectric–photoelastic correction to the dielectric tensor of $\alpha$-quartz introduces an additional anisotropy whose symmetry depends on the direction of $\mathbf{E}^{\mathrm{DC}}$.

\begin{figure}
    \centering
    \includegraphics[width=0.95\linewidth]{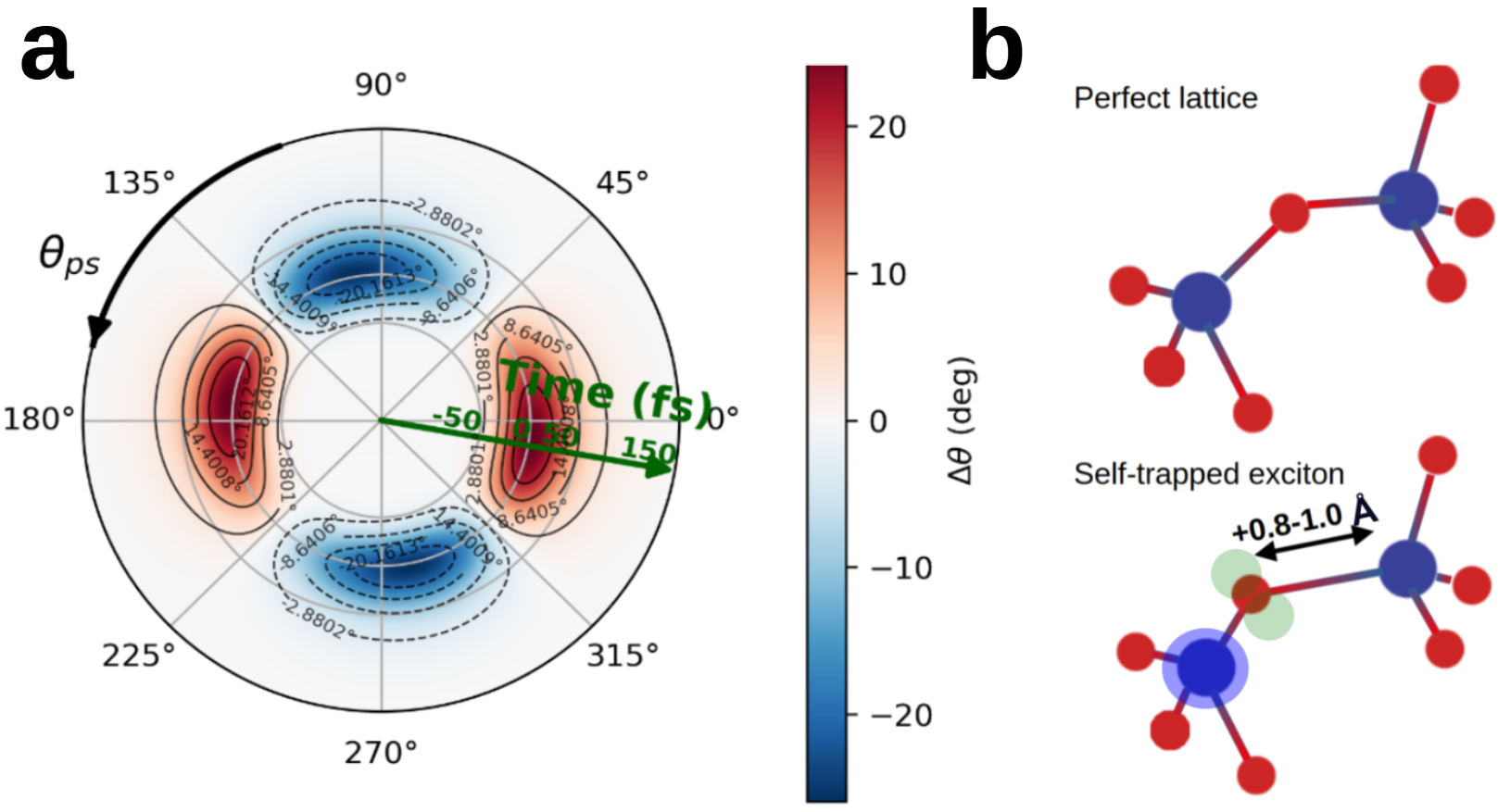}
    \caption{a) Simulated reflectivity $\Delta\theta(t,\theta_{ps})$ map in $\alpha$-quartz. b) Representation of the self-trapped exciton in quartz.}
    \label{fig:fig9}
\end{figure}

In Supplementary Information 5, we present simulations for several orientations of $\mathbf{E}^{\mathrm{DC}}$. When the DC field has equal components along $x$ and $y$, the resulting symmetry does not match our experimental data. However, when $\mathbf{E}^{\mathrm{DC}}$ is taken predominantly along the $x$ direction, we recover the symmetry observed experimentally, as shown in Figure \ref{fig:fig9}a. 

We also remarked that choosing $\mathbf{E}^{\mathrm{DC}}$ preferentially along $y$ yields the correct symmetry but with reversed sign for $\Delta\theta$. Moreover, adding a $z$ component increases the amplitude of the rotation. This can be explained by the non-zero incident angle $\alpha$ of our probe, implying a sensitivity to $z$ component of the permitivity tensor. \\

We simply model its time dependence through a retarded Kerr effect (Eq \ref{eq:retarded_kerr}) even if another behavior is expected. The fluence dependence is also not fathomed as it would recquire ab initio calculation to know how the DC current would diffuse in space with higher excitation rates. Indeed, at higher excitation rates, the high number of carriers could saturating the $x$ channel, and thus a more dispersed plasmon arises. Then a non negligible component of DC field within the $y$ axis reduces the rotation of probe's polarization, acting like a screening.

In conclusion, although a more detailed quantitative analysis would be required to fully determine the relevant parameters, our model qualitatively reproduces the symmetry and the order of magnitude observed in $\alpha$-quartz. This outcome is obtained by assuming the formation of a conduction-electron plasmon oriented along a specific direction in the $(x,y)$ plane. By comparing the two probe orientations in Figure \ref{fig3}, we infer that this crystallographic direction is likely the $[11\bar{2}0]$ axis ($\Gamma$–$K$), which corresponds to the alternating Si–O arrangement shown in Figure \ref{fig3}a and to the location where carriers become trapped into excitonic states \cite{MuozRamo2012}.

The resulting exciton is a self-trapped exciton (STE), with the electron localized near the silicon ion ($3sp$ state) and the hole near the oxygen ($2p$ state). Its formation involves a local distortion of the Si–O bond along $[11\bar{2}0]$, featuring an elongation of about $0.8$–$1.0$\,\AA\ and a $\sim 20^{\circ}$ change of the Si–O–Si angle, as depicted in Figure \ref{fig:fig9}a \cite{MuozRamo2012}. This STE forms within $\sim 150$\,fs after photoexcitation \cite{petite_daguzan_guizard_martin_1996}.

Before this relaxation is complete, a fraction of the conduction electrons remain mobile for $\sim 100$–$150$\,fs along $\Gamma$–$K$, generating a transient DC current. This current induces a strain whose symmetry matches the lattice distortion associated with the eventual STE. Once the structural deformation is established, electrons and holes become trapped in the locally distorted network, forming STEs. Being spatially localized, they no longer sustain a DC current, and the $\Delta\theta$ signal vanishes. Thus, the transient birefringence we observe likely captures the earliest stages of exciton self-trapping in $\alpha$-quartz.

\section{Conclusion}

To conclude, we have systematically investigated the polarization changes induced by a moderately intense femtosecond pulse with simultaneous time and angular resolution, in both transmission and reflection, for amorphous and crystalline ($\alpha$-quartz) $SiO_2$. In transmission, we observe a sizable rotation of the probe polarization (tenths of a degree), while the reflectivity data reveal a striking qualitative difference between amorphous and crystalline samples.

A nonlinear birefringence model reproduces the transmission measurements at low fluence, including both the four-fold symmetry and the amplitude of the rotation. At higher fluence, deviations arise, likely due to a retarded Kerr contribution, higher-order nonlinearities, or propagation effects such as pump depletion or polarization changes during propagation.

For reflection, we developed an anisotropic permittivity model incorporating Pockels and Kerr terms. In amorphous silica—where only Kerr effects are present—the model captures both symmetry and amplitude, although a secondary peak suggests additional mechanisms (carrier-induced absorption or transient surface restructuring).

In $\alpha$-quartz, Pockels and Kerr effect do not reproduce the measurements. Simulations indicate that the conduction electrons drift predominantly along the $[11\bar{2}0]$ direction, generating a transient DC current. This current induces a strain via the inverse piezoelectric effect, facilitating subsequent trapping into self-trapped excitons. Before trapping quenches the DC current, the induced strain perturbs the refractive index, producing the observed polarization rotation.

Finally, we emphasize that significant pump-induced polarization changes occur even well below the damage threshold, as soon as electronic excitation becomes appreciable. These effects depend mainly on the pump–probe relative polarization and can strongly influence time-resolved measurements. Conversely, they may enable ultrafast transient optical components, such as femtosecond waveplates operating in reflection or transmission.

\section*{Funding}
This project received funding from the French Agence Nationale de la Recherche 
(ANR-19-CE30-015-01) under the TOCYDYS program.

\section*{Data Availability}
The datasets generated during and/or analyzed in the present study are available 
from the corresponding author on reasonable request.

\bibliographystyle{unsrt}

\end{document}